\documentstyle{mn2e}
\input epsf
\def\plotone#1{\centering \leavevmode
\epsfxsize=\columnwidth \epsfbox{#1}}

\title[Earth X-ray albedo for CXB radiation]{Earth X-ray albedo for cosmic X-ray background radiation in the
1--1000 keV band}

\author[Churazov,  Sazonov, Sunyaev  and
Revnivtsev]{E.~Churazov$^{1,2}$, S.~Sazonov$^{1,2}$,  R.~Sunyaev$^{1,2}$, 
M.~Revnivtsev$^{1,2}$\\
$^1$ Max-Planck-Institut f\"ur Astrophysik, Karl-Schwarzschild-Strasse 1, 85741
Garching, Germany\\
$^2$ Space Research Institute (IKI), Profsoyuznaya 84/32, Moscow 117997, 
Russia\\
}

\pagerange{\pageref{firstpage}--\pageref{lastpage}}
\pubyear{2001}

\begin{document}
\maketitle

\label{firstpage}
\begin{abstract}
We present calculations of the reflection of the cosmic X-ray
background (CXB) by the Earth's atmosphere in the 1--1000 keV energy
range. The calculations include Compton scattering and X-ray
fluorescent emission and are based on a realistic chemical composition
of the atmosphere. Such calculations are relevant for CXB studies
using the Earth as an obscuring screen (as was recently done by
INTEGRAL). The Earth's reflectivity is further compared with that of
the Sun and the Moon -- the two other objects in the Solar system
subtending a large solid angle on the sky, as needed for CXB studies.
\end{abstract}


%

\sloppypar

\section{Introduction}
Having a mass column density of $\sim 10^3 {\rm g~cm^{-2}}$ at the sea
level the Earth's atmosphere completely blocks the X-rays from celestial
sources. At the same time the outer layers of the Earth's atmosphere reflect
part of the incident X-ray photons due to Compton scattering. The
physical picture is very similar to the well-studied case of the reflection
from a star surface (e.g. Basko, Sunyaev \& Titarchuk, 1974) or an
accretion disk (e.g. George \& Fabian, 1991) except for the different
chemical composition of the reflecting medium. 

The reflection of X-rays by the Earth's atmosphere is important for
e.g. evaluating the echo produced by gamma-ray bursts (Willis et
al., 2005) or for studies of the cosmic X-ray background (CXB). The
present study was particularly initiated by recent observations of
the Earth with the INTEGRAL observatory (Churazov et al., 2007) 
aimed at determining the CXB intensity near the peak of its 
luminosity distribution (i.e. around 30--40 keV). In these observations
the Earth disk blocked the X-rays coming from distant objects, causing a
decrease in the observed flux. The energy dependent reflection of
X-rays by the Earth's atmosphere reduces the amplitude of this
decrement. The purpose of this paper is to provide a simple recipe for
the calculation of the reflected flux in the energy range from few to
few hundred keV.

Since photons change their energy during Compton scattering, the
reflected emission at a given energy depends on the overall shape of
the input spectrum. Generally, one needs to calculate
a Green function describing the reflection of monochromatic
radiation (White, Lightman \& Zdziarski 1988, Poutanen, Nagendra \&
Svensson 1996) and
convolve the incident spectrum with this function to calculate the
reflected spectrum. We instead calculate the energy dependent
``albedo'' $A(E)$ -- the ratio of the flux reflected by the Earth disk
at a given energy to the flux of the incident spectrum at the same
energy, using the canonical approximation of the CXB spectrum (Gruber
et al., 1999) as the incident spectrum. The resulting function $A(E)$
may be used for calculating the reflected emission for an incident
spectrum that has a shape similar to that assumed in our simulations. We
also estimate the uncertainties introduced in $A(E)$ by variations in
the input spectral shape.

Two other bodies in the Solar system subtend a large solid angle
on the sky for telescopes in near-Earth orbits -- the Moon 
and the Sun. We therefore also calculate the X-ray albedo for them and
compare it with the reflectivity of the Earth's atmosphere.

\section{The Earth's atmosphere model}
\label{sec:emod}
According to the standard model of the Earth's atmosphere (see
e.g. \verb$http://www.spenvis.oma.be/spenvis$), a uniform chemical
composition is a reasonably good approximation for altitudes below
$\sim$90 km (the so-called ``homosphere'').  The relative chemical
composition (by volume) of various species in the homosphere is: ${\rm
N_2}$ -- 0.781, ${\rm O_2}$ -- 0.209 and ${\rm Ar}$ -- 0.0093. The
uniformity of the chemical composition is maintained by vertical
winds and turbulent mixing. Above 90 km the chemical composition
starts to vary with altitude, with lighter elements playing an
increasingly more important role. The temperature and the ionization
state of the medium also vary substantially at high altitudes. However
the mass column density of the atmosphere above 90 km is of order
$1.4~10^{-3}$ g and the corresponding Thomson optical depth is only
$\sim 3~10^{-4}$. We therefore choose to neglect the variations of the
chemical composition in the outer layers of the atmosphere and assume
that it is constant throughout the atmosphere.

The entire atmosphere has a very large optical depth at any energy of
interest here (from 1 keV up to $\sim$10 MeV). This ensures that any
characteristic length scale of the problem (e.g. the length scale
corresponding to a unit optical depth at a given energy in the range
from 1 keV to 10 MeV) is of the order of (or smaller than) the scale
height of the atmosphere. All these characteristic length scales are
much smaller than the Earth radius and a plane-parallel atmosphere
should be a reasonably good approximation. The vertical structure of
the atmosphere can then be ignored and the atmosphere can be modeled
as a plane-parallel and uniform slab of matter (see e.g. Mihalas,
1978). In our simulations, the column density of the slab was set to a
large value of $\sim 1000~{\rm g~cm^{-2}}$ so that the slab illuminated from
one side is effectively equivalent to a semi-infinite medium. Further
justification of the plane-parallel slab approximation is given in
section \ref{sec:slab}.

\section{Physical processes}
\label{sec:proc}
The following processes were included in the simulations:
photoelectric absorption, Rayleigh and Compton scattering and
fluorescence.  We completely neglect polarization throughout the
paper (see Poutanen et al. 1996 for the calculation of Compton reflection
from an accretion disk with account for polarization).
An additional process -- electron-positron pair creation
-- takes place for photon energies exceeding $2 \times m_e
c^2=$1022 keV. For our purpose (the atmospheric albedo for CXB
radiation in the 1--1000 keV energy band) the contribution from this
process is very small. This contribution was evaluated using the GEANT
package (Agostinelli et al. 2003) and was found to be less than 1\%
across the 200--1000 keV band, excluding the 511 keV line. Therefore, in the
rest of the paper we neglect this process.

Photoelectric absorption was calculated using the data and
approximations of Verner \& Yakovlev (1995) and Verner et al. (1996). For
fluorescence we use the energies and yields from Kaastra \& Mewe (1993).

\begin{figure}
\plotone{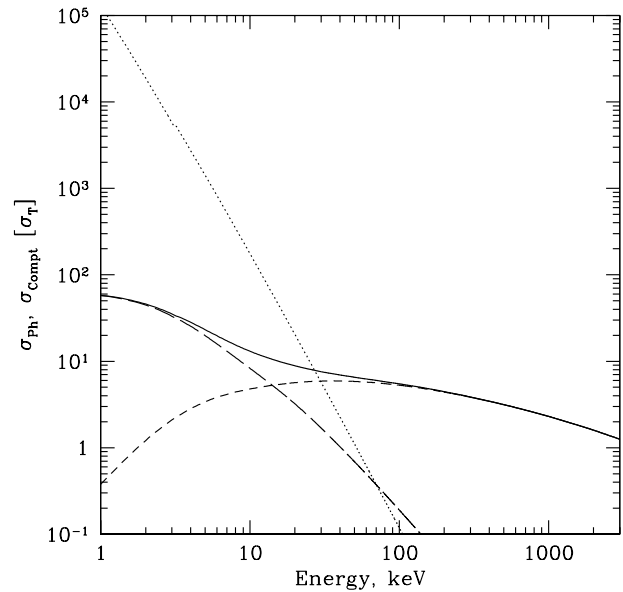}
\caption{Adopted cross sections for photoelectric absorption (dotted
line), Compton (short dash) and Rayleigh (long dash) scattering in
air. The solid line shows the sum of the Compton and Rayleigh scattering
cross sections. All cross sections are given in units of $\sigma_T$
per atom. The nitrogen and oxygen atoms composing ${\rm N_2}$ or ${\rm
O_2}$ molecules are treated as independent atoms. 
\label{fig:cross}
}
\end{figure}

Compton and Rayleigh scattering are the most important processes for
this model. We made use of the same sources of information as are used in the
\verb1GLECS1 package (Kippen 2004) of the GEANT code (GEANT
Collaboration, 2003). Namely the Livermore Evaluated Photon Data
Library (EPDL, see Cullen, Perkins \& Rathkopfand 1990) and the free
electron Klein--Nishina formula are used to calculate total
cross sections and the angular distribution of scattered photons for
each element. The total cross sections for photoelectric, Compton and
Rayleigh scatterings are shown in Fig.~\ref{fig:cross}. Unlike the case
of a typical interstellar medium (ISM), composed of hydrogen, helium
and a small fraction of heavier elements, the Earth's atmosphere is
composed of elements heavier than nitrogen. This difference in
composition has two important consequences. Firstly, the
photoabsorption cross section exceeds the scattering one up
to energies $\sim$30 keV. Secondly, in contrast to the hydrogen
dominated ISM, where the total cross scattering section is constant at low
energies, for the atmospheric composition the Rayleigh scattering
boosts the total scattering cross section at energies below 
10--20 keV (see Fig.~\ref{fig:cross}).

For Compton scattering an additional smearing of the scattered photon
energy due to the distribution of the bound electrons in momentum is
taken into account by using the data on the ``Compton profile'' from Biggs,
Mendelsohn \& Mann (1975). Unlike the case of a usual ISM, where the Compton
profile can produce interesting changes in the spectral shape of the
fluorescent lines (Sunyaev \& Churazov 1996), these effects are less
important for the reflection by air because at the energies of
the most interesting fluorescent lines (e.g. for the $K_\alpha$ line
of argon at 2.96 keV) photoabsorption strongly dominates over Compton
scattering. The CXB itself does not have any sharp features in the
spectrum that could make the effect of smearing on the reflected
spectrum important. Nevertheless, for completeness we include this
effect in the simulations.

When modeling the scattering process, the two nitrogen or oxygen atoms 
composing a ${\rm N_2}$ or ${\rm O_2}$ molecule were regarded as
independent atoms. We are therefore underestimating the cross section for
forward scattering by a factor of 4. The characteristic range of
scattering angles $\theta$ for which this additional increase of the
cross section is important can be estimated from the condition
$2\pi\theta \frac{R}{\lambda}~\ll~1$, where $R$ is the inter-atomic
distance and $\lambda$ is the wavelength of the photon. For a hydrogen
molecule this characteristic angle is $\sim$30--40 degrees for $\sim$6 keV
photons (Sunyaev, Uskov, Churazov 1999) and the increase of the total
cross section is important below 3--4 keV. The inter-atomic distance for
${\rm N_2}$ and ${\rm O_2}$ is $\sim$1.5 $\AA$, i.e. a factor of
two larger than the inter-atomic distance in the hydrogen
molecule. Accordingly the total cross section will change
significantly only at energies as low as $\sim$2--3 keV. As was
mentioned above the albedo at such energies is very low. In the rest
of the paper these effects are neglected.

\section{Model}
\label{sec:model}
When the Earth enters the field of view of an instrument, the Earth
disk blocks the X-rays from distant sources and at the same time the
atmosphere reflects part of the incident X-ray flux. Therefore the net
change of flux  observed by the instrument is the difference between the
CXB flux obscured by the Earth and the reflected CXB flux. To a first
approximation this difference can be expressed as 
\begin{eqnarray}
\Delta F_{\rm obs}(E)=\Omega~ I_{\rm CXB}(E) - \Omega~ \overline{I}_{\rm refl}(E)= \nonumber \\
\Omega ~I_{\rm CXB}(E)~(1-A(E)),
\label{eqn:simple}
\end{eqnarray}
 where $\Omega$ is the solid angle subtended by the Earth, 
  $I_{\rm CXB}(E)$ is the CXB intensity,
  $\overline{I}_{\rm refl}(E)$ is the average intensity of CXB
  radiation reflected from the Earth and $\displaystyle A(E)\equiv
  \frac{\overline{I}_{\rm refl}(E)}{I_{\rm CXB}(E)}\equiv\frac{F_{\rm refl}(E)}{F_{\rm inc}(E)}$ is the albedo of the
  atmosphere. Here $\pi I_{\rm CXB}(E)=F_{\rm inc}(E)$, and the average
  reflected CXB intensity is related to the flux reflected from the
  Earth's atmosphere as $\pi\overline{I}_{\rm 
  refl}(E)\equiv F_{\rm refl}(E)$. Thus an energy dependent factor, $1-A(E)$, relates the
  observed flux and the CXB intensity multiplied by the solid angle
  subtended by the Earth.  Below we calculate $A(E)$ assuming that
the atmosphere can be modeled as a uniform slab of matter. The
validity of this approximation is further addressed in section
\ref{sec:slab}.

We model the reflection by the Earth' atmosphere via the Monte-Carlo
method (see e.g. Pozdnyakov, Sobol \& Sunyaev, 1983).   The
distribution of the input photons over angles follows the $\mu$ law,
where $\mu={\rm cos} \theta $ and $\theta$ is the angle between the
input photon direction and the normal to the surface of the
atmosphere. This distribution corresponds to the case of an element of
the plane surface exposed to isotropic radiation from one side. The
energies of initial photons are sampled according to a given input
intensity $I_0(E)$. The energies and directions of all outgoing
photons are recorded. Of particular interest is the total flux
emerging from the atmosphere (integrated over all angles). Indeed the
same luminosity (flux integrated over the total surface) reflected by
the Earth is going through any imaginary 
surface outside the Earth's atmosphere which completely surrounds the
Earth. This implies that any unit area oriented perpendicular to the
direction towards the Earth center will ``see'' the flux
$\displaystyle F(E)=\frac{F_{\rm refl}(E)4\pi R_{\earth}^2}{4\pi
  D^2}=\overline{I}_{\rm refl}(E)\Omega$, where $D$ is the distance from
the Earth.  Therefore such flux should be seen by any instrument
observing the whole Earth disk at once.

\section{Simulations}
As a starting point we set the shape of the intensity $I_0(E)$ to
be that of the broad-band CXB spectrum in the approximation of Gruber et
al. (1999). Namely:
\begin{eqnarray}
I_{\rm CXB}(E)=\left\{\begin{array}{ll}
7.877~E^{-0.29}~e^{-E/41.13} & 3<E<60~{\rm keV} \\
& \\
0.0259~(E/60)^{-5.5}+ & \\
0.504~(E/60)^{-1.58}+ & E>60~{\rm keV} \\
0.0288~(E/60)^{-1.05} &
\end{array}
\right. 
\label{eqn:cxb}
\end{eqnarray}
Here $I_{\rm CXB}(E)$ is in units of ${\rm keV/keV cm^{-2} s^{-1} sr^{-1}}$.

\subsection{Maximal energy in the input spectrum}
\begin{figure}
\plotone{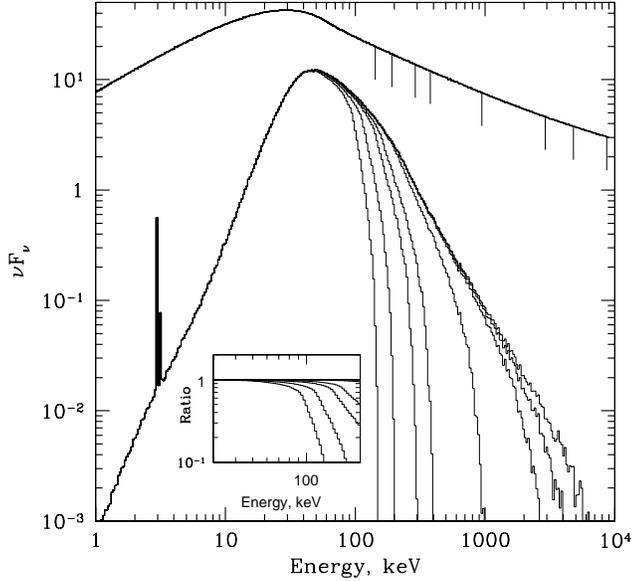}
\caption{Dependence of the reflected intensity ($F_{\rm refl}/\pi$) on the high-energy cutoff $E_{max}$ in the input spectrum. The
spectra shown correspond to $E_{max}=$0.15, 0.2, 0.3, 0.4, 1, 3, 5 and
9 MeV. The inset shows 
the ratio of the spectra with a given $E_{max}$ to that with
$E_{max}=$9 MeV near the maximum of the reflected spectrum. The
top curve shows the input spectra ($I_{\rm CXB}(E)E$), with the cutoff energies
marked with vertical ticks.
\label{fig:emax}
}
\end{figure}

Since we are interested in the reflected spectrum over a broad energy
range (up to $\sim$1 MeV) and at high energies the change of the
photon energy due to Compton recoil is large, it is important to
take into account photons with the initial energy substantially larger
than 1 MeV. Fig.~\ref{fig:emax} shows the effect on the 
reflected spectrum of varying the maximal energy $E_{max}$ in the input
spectrum. The sequence of spectra shown corresponds to $E_{max}$=0.15,
0.2, 0.3 0.4, 1, 3, 5 and 9 MeV. As is clear from this figure, in
order to reproduce the shape of the reflected spectra with a
reasonable accuracy (for illuminating spectra whose shape is not
much different from equation~(\ref{eqn:cxb})) one needs to use a broad energy
range up to at least 5--9 MeV. In the subsequent calculations we use
$E_{max}$=9 MeV. 

The inset in Fig.~\ref{fig:emax} shows a region of the reflected
spectrum near its maximum. It demonstrates that i) the reflected spectrum
near the peak of the CXB spectrum  ($\sim$ 30--50 keV in $\nu F_\nu$
units) is not sensitive to the details of the incident spectrum above
150 keV. Near 100 keV and above, the recoil effect is much more important
and the shape of the reflected spectrum becomes sensitive to the
extrapolation of the incident spectrum to higher energies.

\subsection{Slab approximation}
\label{sec:slab}  As noted in Section \ref{sec:emod}, an assumption
of a plane parallel atmosphere should be a good approximation for the
  problem at hand. We explicitly verified this  in a separate
Monte-Carlo simulation. Compared to the slab model we assumed i)
spherical geometry and ii) an exponential atmosphere with a cutoff
above a certain radius. The same chemical composition and the same set
of physical processes were used as before.  The only difference
between the codes is a computationally more demanding procedure for
the calculation of an optical depth along the path of a photon. In the
most important energy range of 10--200 keV the agreement is excellent
(better than 1\%). The maximum difference of 10\% in the reflected
spectra is seen near 1 MeV, where the albedo $A(E)\ll 1$ and the
factor $(1-A(E))$ is close to unity.

Since we are primarily interested in the angular averaged albedo, we
therefore use the slab approximation for all subsequent
calculations. We note however that if one is specifically interested
in high energy emission emerging at very small angles to the surface,
then the results will depend more strongly on the detailed structure
of the atmosphere's boundary.

The above comparison was done for photons scattered
at least once. In addition there are always photons passing through
the upper layers of the atmosphere without interactions. Since the outer
layers of the Earth's atmosphere may be opaque at low energies and
transparent at high energies, the apparent angular size of the
Earth $\Omega$ (see equation~(\ref{eqn:simple})) does depend on energy. For
instance, an optical depth of unity is reached for a line of sight
having an impact parameter of $\sim R_{\earth}+$ 120 km and $\sim R_{\earth}+$ 70 km
at energies of 1 keV and 1 MeV, respectively.  This effect limits the accuracy of our approximation for a
given $\Omega$  (equation~\ref{eqn:simple}) to $\sim$1--2\%.

\subsection{Dependence on the shape of the input spectrum}
\label{sec:pow}
\begin{figure}
\plotone{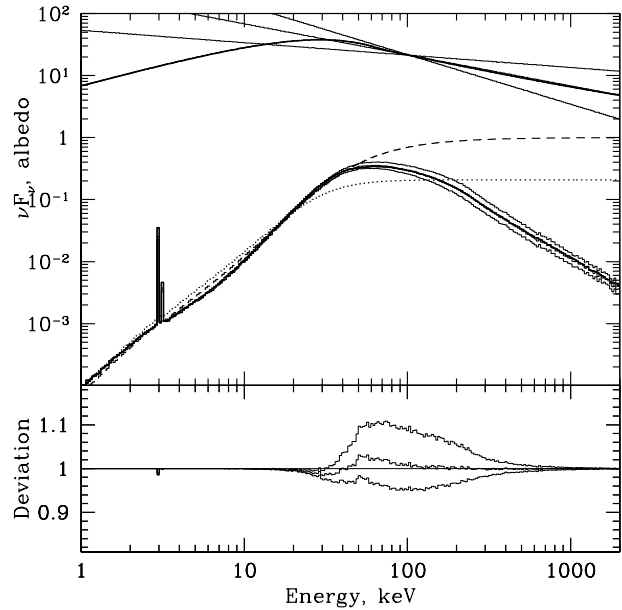}
\caption{ Upper panel: Dependence of the albedo on the shape of
the incident spectrum. The input spectra (shown in the upper part of
the plot) are power laws with photon indices $\Gamma=$~2.2, 2.5 and
2.8. For comparison the CXB spectrum is shown by the thick solid
line. In the lower part of the plot the albedos calculated for these
input spectra are shown. The thick line shows the albedo for the CXB
spectrum. The dotted and dashed lines show the result of a simple
calculation of the albedo from equations (\ref{eq:028}) and
(\ref{eq:two}) respectively. Lower panel: Relative changes in
the factor $(1-A(E))$ calculated assuming a power-law input spectrum
with photon index $\Gamma=$ 2.2, 2.5 and 2.8. 
\label{fig:pow}
}
\end{figure}

Of course the reflected spectrum depends both on the shape and
normalization of the illuminating spectrum. Since we are mainly interested in
the effective atmospheric albedo (i.e. the ratio of the reflected and
input spectra), the dependence on the normalization disappears, but the
shape of the input spectrum still affects the behavior of the albedo
at energies higher than 20--30 keV. Fig.~\ref{fig:pow} shows the
atmospheric albedo for different shapes of the input spectra. The
thick solid lines show our reference CXB input spectrum and the
corresponding albedo. For comparison we show a set of input power law
spectra with photon indices $\Gamma=$ 2.2, 2.5 and 2.8, and
the corresponding albedos.

One can see that below 20--30 keV the shape of the albedo (as a
function of energy) does not depend on the properties of the input
spectrum. This is of course expected since in this regime i)
photoelectric absorption dominates and therefore only first scattering
is important and ii) the change of energy due to recoil is
small. For these energies it is easy to express the
albedo through the ratio of the absorption and scattering cross
sections. Namely, in the single scattering approximation the reflected
flux at energy $E_1$ can be written as:
\begin{eqnarray}
F_{\rm refl,1}(E_1)= \nonumber \\ 
\int{\frac{E_1}{E}I_{\rm CXB}(E)
e^{-\sigma(E)n\frac{z}{\mu}-\sigma(E_1)n\frac{z}{\mu_1}}
~n \sigma_p(E,E_1,\mu_s)} dz dE d\Omega d\Omega_1=\nonumber \\ 
\int{\frac{E_1}{E}I_{\rm CXB}(E)
\frac{\sigma_p(E,E_1,\mu_s)}{\frac{\sigma(E)}{\mu}+\frac{\sigma(E_1)}{\mu_1}}
~ dE d\Omega d\Omega_1},
\label{eqn:gen}
\end{eqnarray}
where $n$ is the particle number density, $\sigma(E)$ is the total
cross section for all processes per particle, $\sigma_p(E,E_1,\mu_s)$
is the differential cross section for the process responsible for
reflected radiation (e.g. Compton scattering or photoelectric
absorption followed by the emission of a fluorescent photon), $z$ is
the vertical coordinate, $\mu$ is the cosine of the angle between the
photon direction and the vertical direction, and $\mu_s$ is the cosine
of the scattering angle. At low energies the recoil effect is weak and
the energy of the photon is conserved: $E_1=E$. One can therefore set
$\sigma_p(E,E_1,\mu_s)=1/2r_e^2(1+\mu_s^2)S(\mu_s,E)\delta(E-E_1)$,
where $r_e$ is the classical electron radius and $S(\mu_s,E)$ is
the form factor. Thus the reflected flux is
\begin{eqnarray}
F_{\rm refl,1}(E)= \nonumber \\
I_{\rm CXB}(E)\frac{1/2r_e^2}{\sigma(E)}\int{S(\mu_s,E)(1+\mu_s^2)\frac{\mu\mu_1}{\mu+\mu_1}
d\Omega d\Omega_1}.
\label{eqn:ref}
\end{eqnarray}
This approximation should work at energies below $\sim$20 keV, where
only first scattering is important. Equation (\ref{eqn:ref})
can be readily integrated for a known form-factor.

For pure Thomson scattering (free and cold electrons),
$S(\mu_s,E)\equiv 1$. For this case, equation~(\ref{eqn:ref}) integrated over
incident angles is explicitly written in Ghisellini, Haardt \& Matt (1994) and
Poutanen et al. (1996). Further integration over $\Omega_1$ yields 
\begin{eqnarray}
F_{\rm refl,1}(E)= 0.654\times I_{\rm CXB}(E)\lambda(E),
\label{eqn:tref}
\end{eqnarray}
where $\lambda(E)=\frac{\sigma_s(E_1)}{\sigma_s(E_1)+\sigma_{\rm ph}(E_1)}$, $\sigma_s(E_1)$ is the total scattering cross section and
$\sigma_s(E_1)+\sigma_{\rm ph}(E_1)$ is the sum of the scattering and
photoabsorption cross sections.  Thus the albedo in a single scattering approximation is
\begin{eqnarray}
A_1(E)=\frac{F_{\rm refl,1}}{\pi I_{\rm CXB}(E)}=0.208\lambda(E).
\label{eq:028}
\end{eqnarray}

This simple approximation works reasonably well up to 20 keV (see
Fig.~\ref{fig:pow}), although the discrepancy of order 50\% is
present, despite that the single scattering approximation is certainly
valid in this regime. The reason for this discrepancy is the large
contribution of coherent/Rayleigh scattering in air to the total
scattering cross section at energies lower than 10--20 keV (see
  Fig.~\ref{fig:cross}). In the case of a multi-electron atom/molecule
  (e.g. nitrogen), the coherent scattering cross section is
  proportional to the square of the number of electrons in the
  system. This makes the phase function at energies $\sim 2$--10 keV very
  elongated in the forward direction (small angle scattering) compared to
  the pure dipole scattering phase function $1+\mu_s^2$. Thus
  equation~(\ref{eq:028}), which was derived assuming the dipole phase
  function,  
  overestimates the reflected flux $F_{\rm refl,1}$. At very low energies
  (of order few keV or less), coherent scattering dominates for
  a wide range of scattering angles and the phase function recovers
  the $1+\mu_s^2$ dependence. Therefore, equation~(\ref{eq:028}) is more
  accurate in this regime. At energies higher than $\sim 10$--20 keV, 
  Compton/incoherent scattering dominates and the phase function again
  approaches the $1+\mu_s^2$ law. Thus equation~(\ref{eq:028}) may work
  well at these energies, although in this regime multiple
  scatterings become important and the single-scattering approximation
  fails (see Fig.~\ref{fig:pow}). 

Instead of the single-scattering approximation one can use the formula
suggested by van de Hulst (1974):
\begin{eqnarray}
A(E)=\frac{(1-0.139~s)(1-s)}{1+1.17~s},
\label{eq:two}
\end{eqnarray}
where $\displaystyle s=\left ( \frac{1-\lambda(E)}{1-g\lambda(E)}
\right )^{1/2}$ and $g=\langle\cos \mu_s\rangle$ is
the asymmetry factor for single scattering (mean cosine of the
scattering angle). For the pure dipole phase
function, $g=0$ and the albedo predicted by equation~(\ref{eq:two}) in the
limit of $\lambda\rightarrow 0$ is $A(E)=0.20\lambda(E)$ -- close  to
the prediction of equation~(\ref{eq:028}). One can calculate 
$g$ as a function of energy with correct account for coherent
scattering to get the best accuracy from equation~(\ref{eq:two}). This
however requires the knowledge of the realistic phase function. We
used instead an energy independent factor $g_{eff}=0.15$ (fit by eye)
and plotted the corresponding curve with the dashed line in
Fig.~\ref{fig:pow}. This approximation works reasonably well (within
10--20\%) up to energies of 30--40 keV.

Figure~\ref{fig:pow} shows that the above simple expressions
(\ref{eq:028}) and (\ref{eq:two}) are accurate to within a factor of two
below 30 keV and can be used for crude estimates.  At higher energies,
full Monte-Carlo simulations are needed to ensure that the accuracy of
even a factor of two is achieved.

At energies higher than 30 keV, the uncertainty in the photon index of
the incident spectrum directly translates into moderate changes in
the albedo. For example, at an energy of $\sim$100 keV, changing the
photon index of the incident spectrum from $\Gamma=2.2$ to 2.8 leads
to a $\sim$25\% change in the albedo. For the particular 
problem of measuring the CXB intensity by Earth occultation the most
important quantity is $(1-A(E))$, where $A(E)$ is the Earth
albedo. This quantity characterizes the modification of the CXB flux
occulted by the Earth due to reflection by the Earth's
atmosphere. Given that the maximal value of the albedo is
$\sim$0.3--0.4, the 25\% changes in the albedo correspond to changes in
$(1-A(E))$ of less than 10\%. This is further illustrated in the
lower panel of Fig.~\ref{fig:pow}, where the ratio
$\displaystyle \frac{1-A_{\rm CXB}(E)}{1-A_{\rm Pow}(E)}$ is shown for photon indices of 2.2,
2.5, 2.8. Here $A_{\rm CXB}(E)$ is the albedo calculated assuming the
shape of the incident spectrum according to equation~(\ref{eqn:cxb}) and
$A_{\rm Pow}(E)$ is the albedo for a power-law input spectrum. It follows from this
figure that an uncertainty of 0.1 in the photon index of
the input spectrum translates into an error of $\sim$2.5\% in the
factor  $(1-A(E))$ around 50--100 keV.

\subsection{Fluorescent lines}
The energies of the fluorescent lines for N, O and Ar are given in
Table 1. All these energies fall in the regime where photoelectric
absorption is the dominant process and only first scattering
matters. The equivalent width of the fluorescent line is the ratio of
the line flux and the scattered continuum. For the line flux we set in
equation~(\ref{eqn:gen}) the cross section
$\sigma_p(E,E_1,\mu_s)=\sigma_{\rm ph,sh}(E)\delta(E_1-E_l)Y_l/4\pi$,
where $\sigma_{\rm ph,sh}(E)$ is the photoabsorption cross section for a
given shell, $E_l$ is the line energy and $Y_l$ is the fluorescent
yield. The line flux is then
\begin{eqnarray}
F_l=
\int{\frac{E_l}{E}I_{\rm CXB}(E)\frac{\sigma_{\rm ph,sh}(E) Y_l }
{4 \pi}\frac{1}{\frac{\sigma}{\mu}+\frac{\sigma_l}{\mu_1}}dE
d\Omega d\Omega_1}= \nonumber \\
\int \frac{E_l}{E} I_{\rm CXB}(E)\sigma_{\rm ph,sh}(E)\frac{Y_l \pi}{3}\times \nonumber\\
\left\{ \frac{1}{\sigma} + \frac{1}{\sigma_l} +  
\frac{\sigma_l}{\sigma^2} 
 \ln\left[ \frac{\sigma_l}{\sigma+\sigma_l} \right] + 
\frac{\sigma}{\sigma_l^2}  \ln\left[ \frac{\sigma}{\sigma+\sigma_l} \right] \right\} dE, 
\label{eqn:line}
\end{eqnarray}
where $\sigma=\sigma(E)$ and $\sigma_l=\sigma(E_l)$. 

The above expression can be readily integrated and the equivalent
width evaluated as the ratio of equations~(\ref{eqn:line}) and
(\ref{eqn:ref}) or (\ref{eqn:tref}).  The expected values of the
equivalent width are given in Table 1. These values are in good
agreement (10\%) with the results of the simulations. The nitrogen and
oxygen line fluxes can be affected also by the coherent scattering off
the entire N$_2$ and O$_2$ molecules, neglected here. We however do not
expect dramatic changes in the equivalent widths of these lines.

\begin{table} 
\caption{Properties of the fluorescent lines.}
\begin{tabular}{l l l l}
\hline
Element & Line energy (keV) & Yield & EW, keV \\
\hline
\hline
N, K$_\alpha$ & 0.39 & 0.0060 & 94.3 \\
O, K$_\alpha$ & 0.52 & 0.0094 & 7.85 \\
Ar, K$_\alpha$ & 2.96 & 0.112 & 2.61 \\
Ar, K$_\beta$ & 3.19 & 0.01 & 0.27 \\
\hline
\hline
\end{tabular}
\label{tab:ew}
\end{table}

\subsection{Angular dependence of the reflected emission}
The angle-dependent Compton reflection by cold electrons has been
considered for instance by Magdziarz \& Zdziarski (1995) and Poutanen et
al. (1996). The reflection from the Earth's atmosphere is qualitatively
similar to that case, especially at high ($>30$ keV) energies. The main
difference is related to the different phase function (because of
coherent scattering) and different chemical composition assumed in the
calculations.

\begin{figure}
\plotone{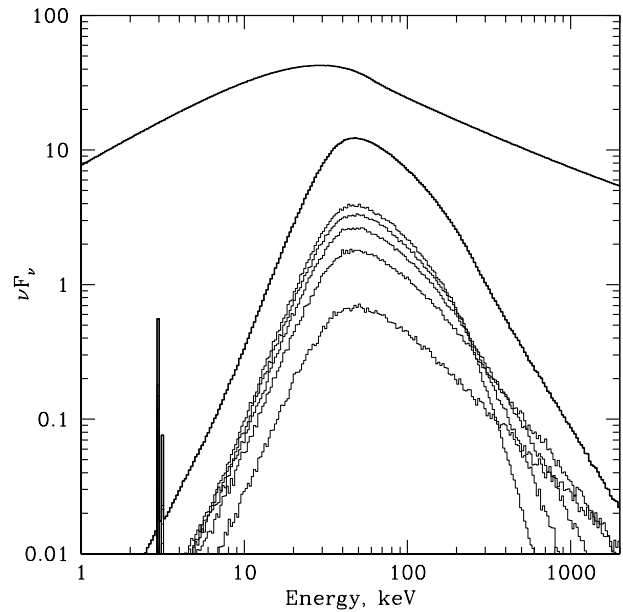}
\caption{Reflected fluxes for several ranges of viewing angle
$\mu=$0.0--0.2, 0.2--0.4, 0.4--0.6, 0.6--0.8 and 0.8--1.0 (thin lines,
from bottom to top at the energies $\sim$30--50 keV), where $\mu$ is the
cosine of the viewing angle with respect to the normal to the
surface. The top two curves (thick lines) show the total incident and
emergent fluxes.
\label{fig:mu}
}
\end{figure}

\begin{figure}
\plotone{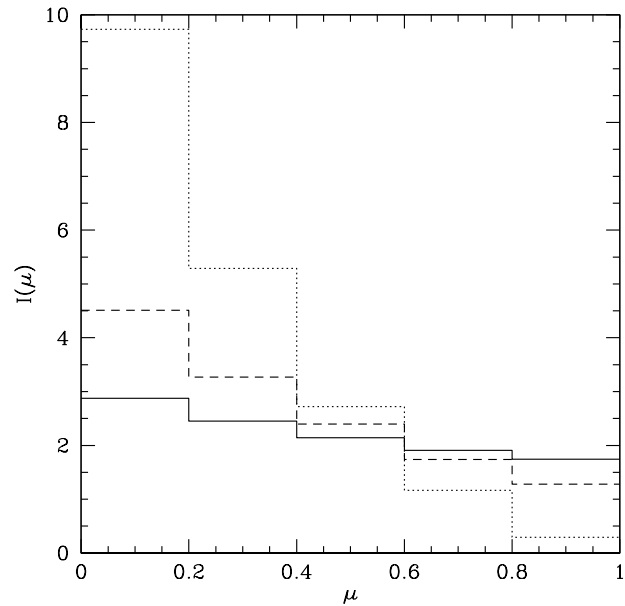}
\caption{Dependence of the flux in the 30--40, 200--300 and 500--600 keV bands
(solid, dashed and dotted curves respectively) on the cosine of
the viewing angle $\mu$. The fluxes are divided by $\mu$ and for a
black-body type angular dependence the curves would be flat.
\label{fig:angular}
}
\end{figure}

The dependence of the reflected emission on the viewing angle is
illustrated in Fig.~\ref{fig:mu}. The thin lines show the reflected
fluxes for the angle ranges $\mu=$0.0--0.2,0.2--0.4, 0.4--0.6,
0.6--0.8 and 0.8--1.0 (thin lines, from bottom to top at the energies
$\sim$30--50 keV), where $\mu$ is the cosine of the viewing angle with
respect to the normal to the surface. For comparison the thick lines
show the total incident and reflected fluxes. One can see that the
shapes the spectra emerging at different viewing angles differ
dramatically. At high energies, the dominant contribution to the
total reflected flux is due to photons emerging at small angles to the
surface. This can be easily understood, since for photons emerging
almost along the normal, the smallest possible scattering angles are
$\sim$90 degrees. This implies a large recoil effect. For instance, for
scattering by 90 degrees the energy of the scattered photon cannot
exceed 511 keV. Thus at energies above 511 keV singly scattered
photons do not contribute to the radiation emerging along the normal
to the surface. This causes a drop in the spectrum at high
energies. On the contrary, for the radiation emerging at small angles
to the surface there is always a contribution from photons scattered
only once (by a small angle). This effect gives rise to a specific
angular dependence of the emerging radiation, especially prominent at
high energies, as demonstrated in Fig.~\ref{fig:angular}. In this
figure we show the dependence of the reflected intensity on the cosine
of the viewing angle $\mu$ for three energy bands: 30--40 keV (solid
line), 200--300 keV (dashed line) and 500--600 keV (dotted line). At
high energies the intensity grows strongly towards small values of
$\mu$ (small angles to the surface). This effect will cause a ``limb
brightening'' for observations of the Earth disk.

We stress that this limb brightening at high energies is
primarily caused by the low reflectivity of the atmosphere for large
scattering angles. The angle averaged albedo is therefore low at
energies where limb brightening is strong. The dependencies shown in
Fig.~\ref{fig:angular} were calculated for a slab geometry and the
exact amplitude of the brightening might change if a more detailed
atmospheric model is used.  However, as was demonstrated in section
\ref{sec:slab}, for the {\it angle averaged albedo} the slab
approximation is sufficiently good.

\subsection{Impact of the chemical composition on the albedo}
The photoelectric cross section drops approximately as $E^{-3}$ above
the energy of the highest absorption edge present in a given compound
(see Fig.~\ref{fig:cross}). The Compton scattering cross section on the
contrary only slowly changes between 10 and 1000 keV. As a result, for
any realistic (for astrophysical conditions) chemical composition of
the reflecting medium, Compton scattering strongly dominates over
photoelectric absorption above 100--200 keV.  This means that 
the shape of the albedo above $\sim$100--200 keV is not sensitive to
the chemical composition of the medium. 

\begin{figure}
\plotone{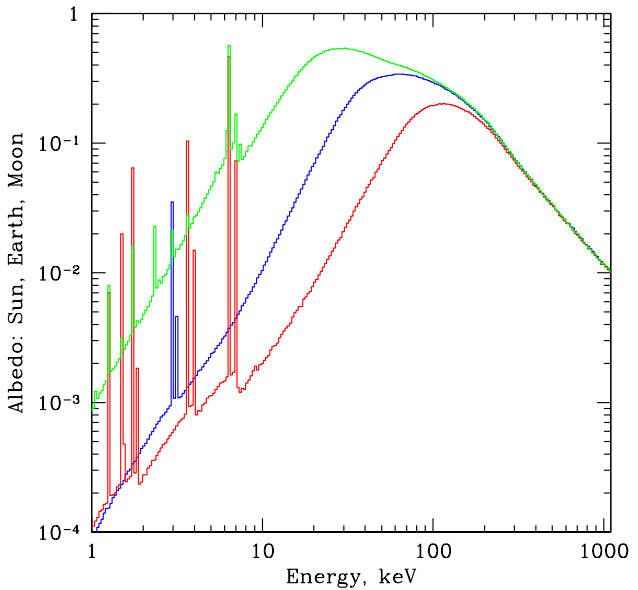}
\caption{Albedo calculated for the CXB spectrum and different chemical 
compositions: solar photosphere (upper curve), Earth's atmosphere
(middle curve) and lunar surface (bottom curve).
\label{fig:moon}
}
\end{figure}

At low energies, on the contrary, photoelectric absorption plays
the dominant role and each material leaves its own imprint on the
reflection albedo. This is illustrated in Fig.~\ref{fig:moon} where the
reflection albedo is shown for three markedly different chemical
compositions. The uppermost curve corresponds to the solar
photospheric chemical composition, i.e. standard hydrogen and
helium dominated gas with a small fraction of heavier elements.  This
``solar'' albedo peaks around 20--30 keV and the reflected spectrum exhibits
a very strong iron fluorescent line at 6.4 keV. The lowest curve was
calculated assuming a chemical composition typical of the Moon's surface --
a mixture of O, Si, Fe, Ca, Al, Mg with a trace of other 
elements. For this chemical composition the photoelectric absorption
plays a much greater role, the reflectivity of the surface at low
energies is much smaller and the peak of the albedo is shifted to
$\sim$100 keV. \footnote{The chemical compositions of highlands
and lowlands of the Moon are substantially different. In particular, the
abundances of iron and aluminium change by a factor of $\sim$2 in the opposite
senses. This causes variations of the albedo at energies
$\sim$30--100 keV by $\sim$20\%. At low energies (below the iron
absorption edge), the fluxes of fluorescent lines and the continuum
also change by factors up to 2. Therefore, the Moon albedo shown in
Fig.~\ref{fig:moon} should be considered less
accurate than those for the Earth and the Sun.} The chemical
composition of the Earth's atmosphere represents an intermediate case
and also the albedo has properties intermediate between the Sun and
the Moon cases.

The albedo is of course smallest in the Moon case, since heavier
elements (compared to the Earth's atmosphere or the solar 
photosphere) dominate the chemical composition. From this point of
view the Moon is a better screen for the CXB than the
Earth. For a typical satellite orbit (in particular for INTEGRAL),
the angular size of the Moon is $\sim$30' arcmin only (diameter) and
obscuration by the Earth has a very strong advantage in terms of
the subtended solid angle.

Both the Moon and the Sun could produce reflected signals associated
with bright gamma-ray bursts or other transient events. For instance, recently
the Helicon instrument on board the Coronas-F spacecraft detected a
giant outburst from the soft gamma-ray repeater SGR~1806$-$20 reflected
by the Moon (Mazets et al., 2005, Frederiks et al., 2007). In this
particular observation the repeater and the Sun were both occulted by
the Earth and only the emission reflected by the Moon was
detected. If the mutual orientation of the objects had been different,
the Sun could have produced a stronger (especially at energies below 100
keV) reflection signal.

\section{Discussion}
\begin{figure}
\plotone{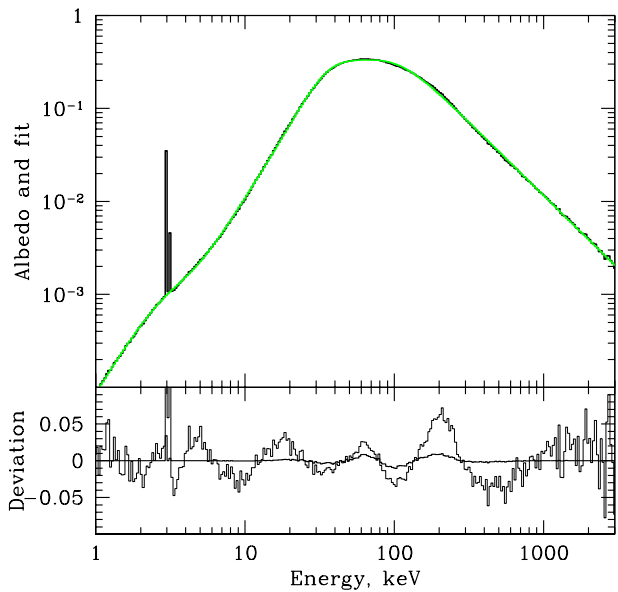}
\caption{Comparison of the albedo $A(E)$ integrated over all angles
(thick gray curve) and the analytic approximation (black curve). The
thin curve in the lower panel shows the relative deviation of the
simulated albedo from the analytic approximation. For the CXB studies
more important is the absolute value of the difference between the
simulated albedo and the analytic approximation. This difference is
shown in the lower panel with the thick solid line. It is less than
1\% at any energy in the 1-1000 keV range, except at the energy of
the argon fluorescent line.
\label{fig:analytic}
}
\end{figure}


The energy dependence of the angle averaged Earth albedo (evaluated
for the input spectrum given by equation~(\ref{eqn:cxb})) can be approximated in the
1--1000 keV energy range by the following formula (see Fig.\ref{fig:analytic}):
\begin{eqnarray}
A(E)=\frac{1.22}
{\left(\frac{E}{28.5}\right)^{-2.54}+\left(\frac{E}{51.3}\right)^{1.57}-0.37}\times
\nonumber \\
\frac{2.93+\left(\frac{E}{3.08}\right)^4}{1+\left(\frac{E}{3.08}\right)^4}\times
\nonumber \\
\frac{0.123+\left(\frac{E}{91.83}\right)^{3.44}}{1+\left(\frac{E}{91.83}\right)^{3.44}}.
\end{eqnarray}
The factor $\Omega~(1-A(E))$ was used by Churazov et al. (2007) to
relate the CXB spectrum and the decrease of hard X-ray flux observed
by INTEGRAL during Earth observations. This factor provides a
convenient way of correcting for the atmosphere reflection (for a
given shape of the incident spectrum) as long as the full Earth disk
is observed and the variations of the telescope response across
the disk can be ignored (see section \ref{sec:model}). In particular,
it does not depend on the distance from the Earth. This universality
breaks down if only part of the Earth disk is observed.

As is shown in Section \ref{sec:pow}, the angle averaged albedo
exhibits a modest dependence on the assumed shape of the input
spectrum. The uncertainty of 0.1 in the photon index of the input
spectrum used for the albedo calculation translates into an error of
$\sim$2.5\% in the derived CXB flux at $\sim$50--100 keV. An
additional uncertainty of $<$2\% is associated with the slab
approximation (see Section \ref{sec:slab}). As mentioned in
Section \ref{sec:proc}, we did not model at all the polarization of the
radiation. The incident isotropic radiation is unpolarized and the
total flux reflected from the full Earth disk is of course also
unpolarized. We also verified in explicit Monte-Carlo simulations that
in calculating the spherical albedo, the substitution of the full Rayleigh
scattering matrix by the Rayleigh phase function does not lead to albedo
changes exceeding 1\% for any value of the single scattering albedo $\lambda$.

At high energies (above $\sim$50--100 keV) the Earth's atmosphere
becomes a powerful source of hard radiation induced by the interaction
of cosmic rays with the atmosphere. Typical spectra of the atmospheric
emission are calculated in Sazonov et al. (2007). The reflection of
the CXB photons in this regime is of secondary importance.

\begin{figure}
\plotone{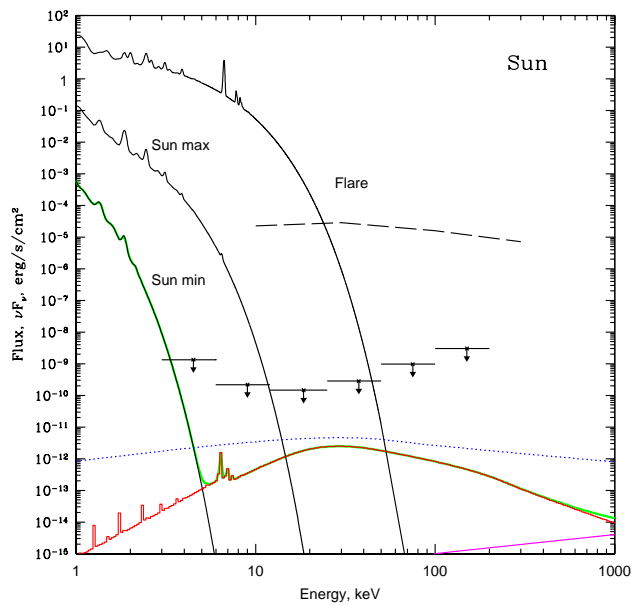}
\caption{Reflected CXB emission (red line) from the Sun disk in
comparison with other solar emission components. The three (soft) thermal
spectra (solid black lines) are models of typical solar corona
emission during solar minimum, maximum and flares (from Peres et al.
2000). The nonthermal component of a strong solar flare is shown with
the dashed line. Upper limits show the constraints on the quiet Sun
emission obtained by RHESSI (Hannah et al. 2007). The solid magenta line
(bottom-right part of the plot) shows the emission induced by the
interactions of cosmic rays with the solar atmosphere (Seckel et al. 1991,
Sazonov et al. 2007). The thick solid green line shows the expected
spectrum for the most quiet Sun. The CXB spectrum integrated over the
solid angle subtended by the Sun disk is also shown for comparison
(blue dotted line).
\label{fig:suna}
}
\end{figure}

It is interesting to compare the reflected CXB emission from the
Earth, Moon and Sun in the 1--1000 keV energy range to other
components of their X-ray spectra (see e.g. Bhardwaj et al. 2007 for a
review of X-ray properties of Solar system objects, especially in soft
X-rays). Since we are concentrating on the energies above 1 keV, the
main components are: i) direct or reflected solar radiation (dayside
for planets), ii) reflected CXB emission and iii) emission induced by
the interaction of cosmic rays with the atmospheres/surfaces of these
objects . In Fig.~\ref{fig:suna} we show typical X-ray spectra (black
solid lines) of the Sun during solar minimum, maximum and a flare,
using the emission measures and gas temperatures from Peres et
al. (2000). The dashed black line schematically shows the nonthermal
emission component of a powerful solar flare, while the upper limits
illustrate the constraints on the quiet Sun emission by RHESSI (Hannah
et al. 2007). One can see that the reflected CXB emission (red solid
line) is below the solar X-ray emission at energies below 5 keV even
during the solar minimum. At higher energies the reflected component
is more than two orders of magnitude fainter than the RHESSI upper
limits, but potentially could be an important component in the Sun
X-ray emission at these energies. The emission induced by cosmic-ray
interactions with the solar atmosphere (Seckel et al. 1991, Sazonov et
al. 2007) is shown by the magenta line. This component is weak
compared to the reflected CXB radiation up to the highest energies of
interest here (up to 1 MeV).

\begin{figure}
\plotone{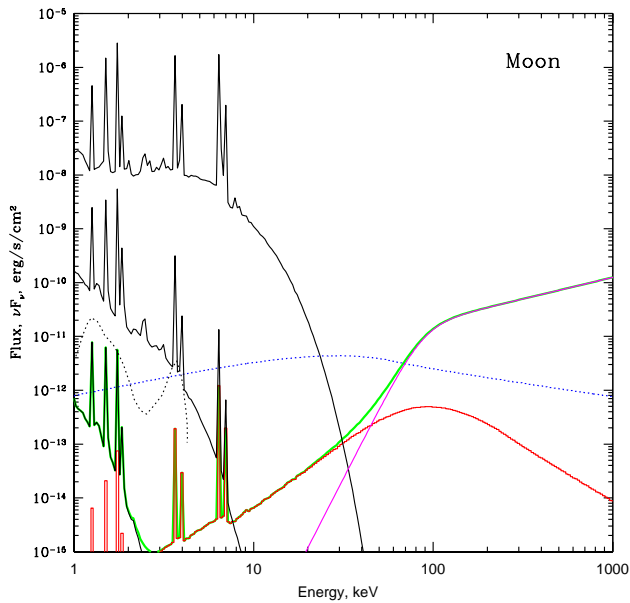}
\caption{The same as in the previous figure, but for the Moon. The three
spectra shown by black solid lines are the expected
solar emission reflected from the dayside of the Moon for solar minimum,
maximum and a flare (without the nonthermal component) 
respectively. The black dotted line schematically shows the recent
SMART-1/D-CIXS measurements (Grande et al. 2007). The green thick
solid line shows the total expected dayside emission from the Moon.  
\label{fig:moona}
}
\end{figure}

For the Moon (Fig.~\ref{fig:moona}), the picture is quite
different. The key feature is the presence of numerous fluorescent
lines and small reflectivity in the continuum. In this figure we show
the solar emission (again for solar minimum, maximum and a flare; black
lines) reflected by the day-side of the Moon, along with the reflected
CXB emission (red) and the cosmic-ray-induced component (magenta). The
reflection of the solar radiation was observed from the lunar orbit
with Luna 12 (Mandel'shtam et al., 1968), Apollo 15 and 16 (e.g. Adler
et al., 1973) and more recently with SMART-1/D-CIXS (Grande et
al. 2007). The black dotted line in Fig.~\ref{fig:moona} schematically
shows the SMART-1/D-CIXS measurements, crudely converted to the units
used in our plots from the published raw count spectra. The Moon
albedo with respect to the CXB is smallest (compared to the Sun and
the Earth), while the cosmic-ray induced emission is highest, since
the Moon has no magnetic field, which for the Earth (and especially
for the Sun) introduces a low-energy cutoff in the distribution of
cosmic particles capable to reach the atmosphere/surface. The
reflected CXB emission and the cosmic-ray induced component intersect
near 40--50 keV. Below this energy (but above 1 keV) the dark side of
the Moon is expected to be very X-ray dark (as indeed observed,
Wargelin et al., 2004) -- at the level of a few per cent of the CXB
surface brightness, except for the fluorescent lines.

\begin{figure}
\plotone{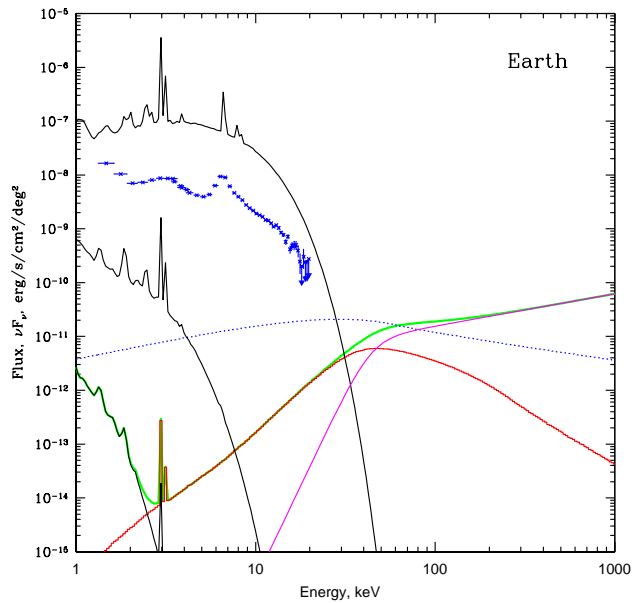}
\caption{The same as in the previous figure, but for 1 square degree
of the Earth's atmosphere.  Data points (blue asterisks) shows the
reflected spectrum of the Earth dayside measured by RXTE during strong
Solar flare (Molkov et al., private communications).
\label{fig:eartha}
}
\end{figure}

The Earth CXB albedo along with the reflected Solar radiation
from the day-side, and cosmic ray induced component are shown in
Fig.\ref{fig:eartha}. For comparison the reflected spectrum of the
Earth day-side measured by RXTE during strong Solar flare (blue data
points, Molkov et al., in preparation) is shown. The surface
brightness of the X-ray aurora may have similar levels at energies
below few tens of keV.

In terms of the albedo strength and the surface brightness of the cosmic ray
induced emission the Earth case is intermediate
between the Sun and the Moon cases (see also Fig.\ref{fig:moon}). The
``advantage'' of the Earth is the large solid angle subtended by the
Earth disk, which makes the signal larger for the wide-field
instruments. The spectrum measured by INTEGRAL (Churazov et al., 2007)
in the 5-200 keV range is a combination of the obscured CXB emission,
reflected CXB emission and the atmospheric emission. The results were
found to be consistent with expectations for the canonical CXB spectral
shape and theoretical calculations of the albedo (this work) and the
atmospheric emission (Sazonov et al., 2007). For INTEGRAL observations
the albedo was making important correction in the energy 
range of 20-100 keV (see Fig. 10 in Churazov et al., 2007). At
energies above 60-70 keV the signal measured by INTEGRAL was dominated
by the cosmic ray induced component with the observed flux very close
to the GEANT calculations by Sazonov et al., 2007. As pointed out in
Churazov et al., 2007 and Sazonov et al., 2007 this overall agreement
of the observed and predicted spectra strongly suggests that the Earth
can be used as a useful calibrator for future hard X-ray and gamma-ray
missions.

\section{Conclusions}
We calculated the Earth atmospheric albedo for the CXB radiation in
the 1--1000 keV energy range. An analytic approximation for the angle
averaged albedo is provided, which is especially useful when the whole
Earth disk is observed. These calculations (along with the
calculations of the cosmic ray induced atmospheric emission by Sazonov
et al., 2007) were used in the analysis of the INTEGRAL observations
of the Earth (Churazov et al., 2007) and a good agreement between
predictions and measurements was found.

We further compared the Earth albedo to the CXB albedo for the Sun and
the Moon and discuss other components contributing to the X-ray
emission of these objects in the 1-1000 keV band.

The night-side of the Moon should be the X-ray darkest object in the
solar system (subtending substantial solid angle for the telescope at
the Earth orbit) in the energy range 1-30 keV, except at the energies
of fluorescent lines of (e.g. Si or Fe).

The reflectivity of the Sun is on the contrary the highest and for
exceptionally strong gamma-ray bursts or other transient events the
signal reflected by the Sun below 100 keV (at the level of few
$10^{-6}$ of the direct signal) will be stronger than by the Moon (see
Mazets et al., 2005, Frederiks et al., 2007 for the discussion of a
recent outburst from SGR1806-20 reflected by the Moon).

The Earth, because of the large solid angle subtended by its disk (for
a wide-field telescope on the Earth orbit) is the most useful object
for the CXB obscuration studies and potentially for the flux
calibration of hard X-ray and gamma-ray missions at energies higher
than 50-100 keV.

\section{Acknowledgements} 
We are grateful to the referee, Juri Poutanen, for many useful
comments and suggestions. This work was supported by the DFG grant
CH389/3-2 and the program of the Russian Academy of Sciences "Origin
and evolution of stars and galaxies''.

\label{lastpage}
\end{document}